\documentstyle[preprint,aps,epsf]{revtex}
\begin{document}

\draft

\title{Comment on: Dynamic and Static Properties of the\\
Randomly Pinned Flux Array}

\author{Heiko Rieger}
\address{Institut f\"ur Theoretische Physik, Universit\"at zu K\"oln,
50926 K\"oln, Germany}
\address{HLRZ c/o Forschungszentrum J\"ulich, Postfach 1913,
52425 J\"ulich, Germany}

\date{February 13, 1995}

\maketitle

\begin{center}
\parbox{13cm}{
\baselineskip12pt
  We reconsider the model of the randomly pinned planar flux array (or
  the two-dimensional XY-model with random fields and no vortices)
  in extensive Monte Carlo simulations. We show that contrary to what
  has been reported recently [Phys.\ Rev.\ Lett.\
  {\bf 72}, 4133 (1994)] the disorder-averaged displacement
  correlation function is different from the pure case for larger
  values of the disorder. Its slope as a function of the logarithm
  of disorder is compatible with newer RG-calculation results.
}
\end{center}

\pacs{}

\input{psfig}

In a recent letter Batrouni and Hwa \cite{lett} reported results
of numerical simulations of the planar flux array described by
the random phase model
\begin{equation}
H= \frac{\kappa}{2} \sum_{\langle ij\rangle} (u_i-u_j)^2
- \lambda\sum_i\cos(2\pi u_i - \beta_i)\;,
\label{ham}
\end{equation}
where $\langle ij\rangle$ denote nearest neighbor sites on a square
lattice, $u_i$ a real-valued displacement-like field, $\beta_i$ a
random phase uniformly distributed in the interval $[0,2\pi]$, and
$\lambda$ the strength of the pinning potential. The main conclusion
of the investigation in \cite{lett} was {\bf a)} that the various
renormalization group predictions existing up to that time could be
ruled out by their numerical results, which I agree with,
and {\bf b)} that the disorder averaged correlation function
$C(r)=[\langle (u_i-u_{i+{\bf r}})^2\rangle]_{\rm av}$,
where $\langle\cdots\rangle$ means the thermodynamic expectation
value and $[\cdots]_{\rm av}$ the disorder average,
is indistinguishable from the pure case, i.e.\
$C(r)=C_{\lambda=0}(r)=T/\kappa\pi\log r$ for
$T\le T_g=\kappa/\pi$.

In this comment I would like to point out that the last statement is a
consequence of the weakness of the disorder they used, namely
$\lambda=0.15$, and that for stronger disorder (or larger length
scales) the correlation function $C(r{\bf})$ differs significantly
from the pure case. Furthermore the numerical data I obtain are
compatible with the analytic predictions of \cite{Korsh,Giam} and the
recent numerical results of a related model \cite{Shapir} with
infinite disorder.

I used the usual Monte Carlo algorithm to calculate the static
expectation values of the model (\ref{ham}). By setting $\kappa=2$ the
glass transition temperature is $T_g=2/\pi\approx0.637$.  The system
sizes were L=32 and 64 (where 1280 and 256 samples were used,
respectively) and periodic boundary conditions were imposed, which
forces the correlation function in the $x$-(and $y$)-direction to be
symmetric $C(r)=C(L-r)$, $r\le L/2$. By applying a least square fit of
the data points for $C(r)$ with $r\le L/4$ (where the effect of the
periodic boundary conditions is still negligible) to the function
$C(r)=a+b(T)\log r$ ($T$ being the temperature) one gets the results
shown in figure 1 for different disorder strengths.

For weak disorder $\lambda\sim0.15$ the slope $b(T)$ is indeed
indistiguishable from the pure case $b(T)=T/\kappa\pi$, as Batrouni and Hwa
observed \cite{lett}. However, by increasing the disorder
($\lambda\ge0.5$) one obtains a slope $b(T)$ that is significantly
different from the pure case already in the vicinity of $T_g$.  Since
systems with larger disorder are hard to equilibrate, only data for
$\lambda\le2$ are shown, but the trend seems to be obvious: the
estimate of $b(T)$ obtained from intermediate length scales increases
with increasing disorder strength.

One should point out that $b(T)$ shown in
figure 1 can only be interpreted as the slope of $C(r)$ as a function
of $\log r$ for small length scales.  For certain values of $T$ and
$\lambda$ one observes a significant upward bend in these curves
already at length scale that smaller than $L/4$, indicating that the
asymptotic behavior is not yet reached.  In figure 2 the derivative of
$C(r)$ with respect to $\log r$, thus yielding a local slope $b(T,r)$,
is depicted for $\lambda=0.5$ and $T=0.4$. This plot gives further
support for the hypothesis that $\lim_{r\to\infty}b(T,r)=T_g/2\pi$,
however, a quadratic dependency $C(r)\sim\log^2 r$, implying
$\lim_{r\to\infty}b(T,r)=\infty$, cannot be excluded. With this
observation in mind the data shown in figure 1 have to be interpreted
as {\it lower bounds} for the asymptotic slope $b(T)$ of $C(r)$ as function
of $\log r$. Hence the results
for the correlation function are compatible with $b(T)=T_g/\kappa\pi$ for
$T\le T_g$, however, also a quadratic dependency $C(r)\sim\log^2 r$,
implying $\lim_{r\to\infty}b(T,r)=\infty$, although unlikely,
cannot be strictly excluded.

Concluding I have presented numerical evidence that the disorder
averaged spatial correlation function of model (\ref{ham}) is indeed
distinct from the pure case $\lambda=0$. For weak disorder this
becomes manifest only on length scales that are not attainable via
Monte Carlo simulations yet, which is the reason why Batrouni and Hwa
\cite{lett} were not able to detect it. However, my results
agree with their conclusion that the various RG prdictions, prior
to their work, were incorrect.

I would like to thank T.\ Nattermann, S.\ Scheidl, L.\ H.\ Tang and
J.\ Kierfeld for many valuable suggestions and discussions.
The computations were done on the Parsytec--GCel1024 from the Center
of Parallel Computing (ZPR) in K\"oln and the Intel Paragon System
from the Supercomputercenter (HLRZ) at the Forschungszentrum J\"ulich.
This work was performed within the SFB 341 K\"oln--Aachen--J\"ulich.

\begin{figure}
\epsfxsize=13cm
\epsffile{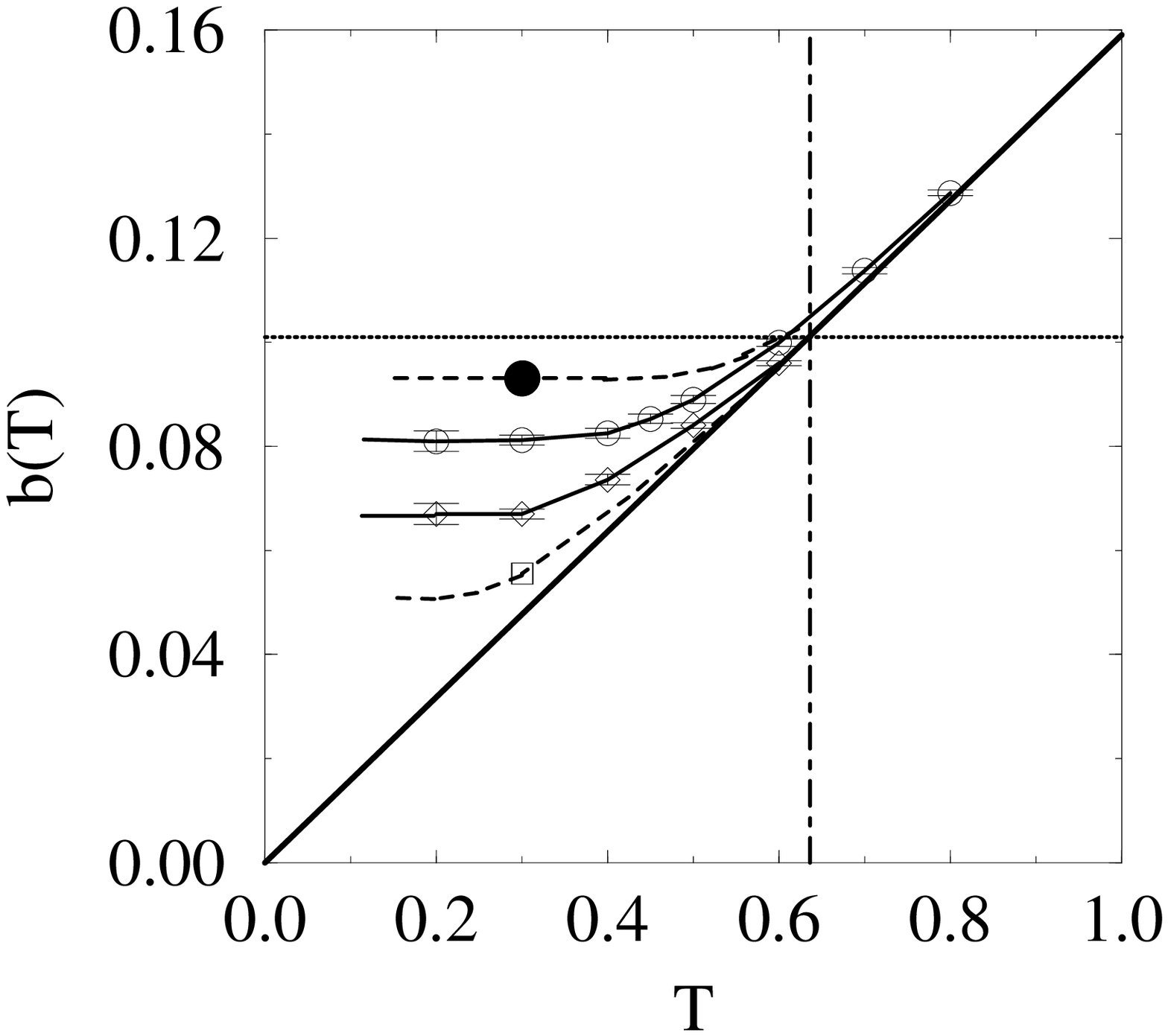}
\vskip0.9cm
\caption{ \label{fig1}
The slope $b(T)$, obtained via a least square fit of the
MC-data for $r\le L/4$, as a function of the temperature $T$. From top
to bottom one has $\lambda=2.0$ ($\bullet$), $1.0$
($\protect{\circ}$), $0.5$ ($\diamond$) and $0.25$
($\protect{\Box}$). Data for $\lambda=0.15$, as obtained by Batrouni
and Hwa [1] fall onto the thick full line, which is the analytic
result for the pure model: $b(T)=T/\kappa\pi$. The dash-dotted line
indicates the glass transition temperature $T_g=\kappa/\pi$ and the dotted
line is at $b(T)=T_g/\kappa\pi$.}
\end{figure}

\begin{figure}
\epsfxsize=13cm
\epsffile{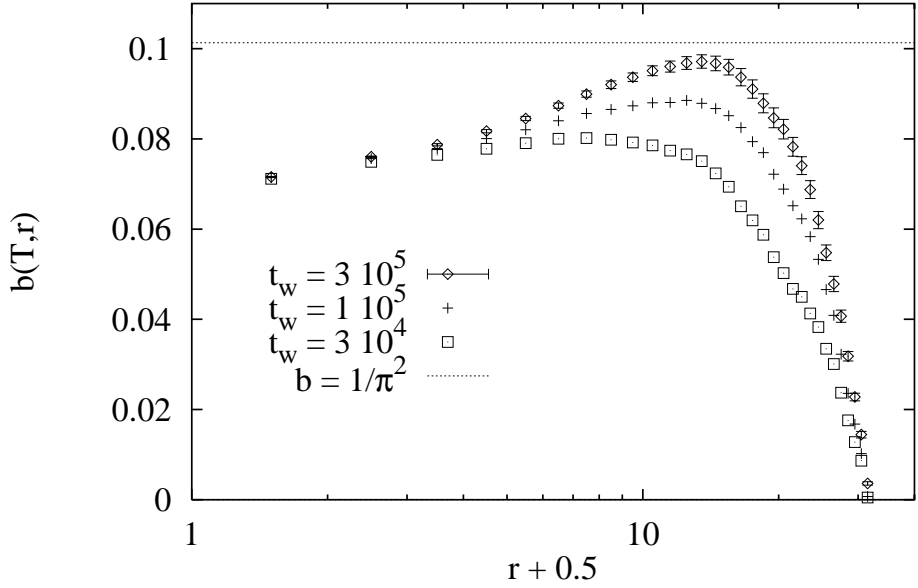}
\vskip0.9cm
\caption{ \label{fig2}
The local slope $b(T,r)=[C(r+1)-C(r)]/[log(r+1)-log(r)]$ as a function
of distance. Shown are data for three different equilibration times:
$t_w=3\cdot10^4$ ($\Box$), $t_w=10^5$ (+) and $t_w=3\cdot10^5$ ($\diamond$).
Curves for larger equilibration times are identical to the last one
meaning that equilibration has been achieved. The dotted line is at
$b=T_g/(\pi\kappa)=1/\pi^2$. Because of the periodic
boundary conditions of the finite system of length $L$ one has
$b(T,L/2)=0$.  If the prediction
$C({\bf r})=T_g/(\pi\kappa)\log\vert{\bf r}\vert$ for $T\le T_g$ holds,
$b(T,r)$ has to approach this line for $r\to\infty$ in the infinite system.}
\end{figure}

\end{document}